\documentclass{article}

\usepackage{microtype}
\usepackage{graphicx}
\usepackage{subfigure}
\usepackage{booktabs} 
\usepackage{tablefootnote}
\usepackage{hyperref}

\usepackage[accepted]{icml2024}

\usepackage{amsmath}
\usepackage{amssymb}
\usepackage{mathtools}
\usepackage{amsthm}
\usepackage{comment}
\usepackage[flushleft]{threeparttable}
\usepackage[textsize=tiny]{todonotes}
\usepackage{caption}

\usepackage[capitalize,noabbrev]{cleveref}

\theoremstyle{plain}
\newtheorem{theorem}{Theorem}[section]
\newtheorem{proposition}[theorem]{Proposition}

\theoremstyle{definition}
\newtheorem{definition}[theorem]{Definition}

\theoremstyle{remark}
\newtheorem{remark}[theorem]{Remark}

\usepackage[textsize=tiny]{todonotes}
\usepackage{diagbox}
\usepackage{bm}
\usepackage[free-standing-units=true]{siunitx}
\usepackage[font=small,labelfont=bf]{caption}

\icmltitlerunning{Combining Neural Networks and Symbolic Regression for Analytical Lyapunov Function Discovery}

\begin{document}

\twocolumn[
\icmltitle{Combining Neural Networks and Symbolic Regression for Analytical Lyapunov Function Discovery}



\icmlsetsymbol{equal}{*}

\begin{icmlauthorlist}
\icmlauthor{Jie Feng}{equal,yyy}
\icmlauthor{Haohan Zou}{equal,yyy}
\icmlauthor{Yuanyuan Shi}{yyy}
\end{icmlauthorlist}

\icmlaffiliation{yyy}{Department of Electrical and Computer Engineer, UC San Diego, La Jolla, United States}

\icmlcorrespondingauthor{Jie Feng}{jif005@ucsd.edu}

\icmlkeywords{Machine Learning, ICML}

\vskip 0.3in
]



\printAffiliationsAndNotice{\icmlEqualContribution} 

\begin{abstract}
We propose CoNSAL (Combining Neural networks and Symbolic regression for Analytical Lyapunov function) to construct analytical Lyapunov functions for nonlinear dynamic systems. This framework contains a neural Lyapunov function and a symbolic regression component, where symbolic regression is applied to distill the neural network to precise analytical forms. Our approach utilizes symbolic regression not only as a tool for translation but also as a means to uncover counterexamples. This procedure terminates when no counterexamples are found in the analytical formulation.
Compared with previous results, CoNSAL directly produces an analytical form of the Lyapunov function with improved interpretability in both the learning process and the final results. We apply CoNSAL to 2-D inverted pendulum, path following, Van Der Pol Oscillator, 3-D trig dynamics, 4-D rotating wheel pendulum, 6-D 3-bus power system, and demonstrate that our algorithm successfully finds their valid Lyapunov functions. Code examples are available at \href{https://github.com/HaohanZou/CoNSAL}{github}.
\end{abstract}

\section{Introduction}
\label{intro}
The field of deep learning has sparked significant interest in learning and data-driven control techniques for nonlinear systems. Yet, a major hurdle for the practical implementation of learning-based methods is their lack of guaranteed stability or safety, alongside interoperability issues \cite{amodei2016concrete}. To address this challenge, a common solution involves identifying valid certification functions for nonlinear dynamical systems. A well-known certificate function for stability guarantees is the Lyapunov function \cite{khalil2002nonlinear}, which is an energy-based function used to prove the stability of the equilibrium point. 
The Lyapunov functions are indispensable for control system designers, enabling to affirm the stability of complex dynamics and offer vital insights into system behaviors. The significance of such functions has led to the development of various computational construction methods \cite{Peter}, where \cite{mcgough2010symbolic} deploys an evolutionary algorithm for symbolic computation of Lyapunov functions and \cite{sos,dai2023convex} use sum-of-squares (SOS) methods. However, these methods either require pre-defined function templates or lack flexibility due to limited candidates. 

Recent advancements in learning-based methods have paved the way for successful data-driven techniques in the discovery of neural-network-based Lyapunov functions, as demonstrated in studies such as \cite{edwards2024fossil, wang2024lyapunov, yang2024lyapunovstable, wu2023neural, zhou2022neural, chang2019neural}. We refer readers to a recent review \cite{10015199} for more information. Despite the advancements, current approaches face two major challenges: 1) Generalization and 2) Scalable Verification \cite{10015199}. The issue of generalization arises from the fact that these learned certificates are trained on specific dynamics within a specific region on fixed parameters. As a result, it is not trivial to generalize the learned Lyapunov functions to a greater state space or adapt to slightly different dynamics. Furthermore, an effective solution for verifying certificates in networks remains elusive, as the existing formal verification methods including satisfiability modulo theories (SMT) \cite{chang2019neural}, mixed-integer linear programming (MIP) \cite{wu2023neural}, and $\alpha,\beta$-CROWN \cite{yang2024lyapunovstable} can be computationally expensive, which restricts their applicability in complex systems.

On the contrary, analytical Lyapunov functions have two benefits: 1) It is interpretable and can potentially guarantee global stability, which enhances its generalizability and may provide further insights to scientists. Specifically, a stability-guaranteed control policy can be designed with a known analytical Lyapunov function \cite{10163934,10336939,cui2023structured}. 2) It can bypass the neural network verification. Building on the preceding discussion, our research focuses on the following question:
\emph{Can method based on Neural Networks directly yields analytical Lyapunov functions for nonlinear dynamical systems?} 

To tackle this challenge, we propose CoNSAL, short for Combining Neural networks and Symbolic regression \cite{angelis2023artificial} for Analytical Lyapunov function, which comprises a neural network learner and a falsifier similar to \cite{chang2019neural}. Unlike this method, which utilizes SMT solvers for validation, we employ symbolic regression to elucidate the neural Lyapunov function in analytical form and verify stability conditions with its roots. Stability is ascertained if the nonnegative Lyapunov function's only root in a specified region corresponds to the desired equilibrium point (such as the origin), and its nonpositive Lie derivative either has no roots or a single root at its equilibrium. Global stability can be concluded when the conditions are globally valid. If the stability condition is not met, we efficiently generate counterexamples near the root, adding them to the training set to update the neural Lyapunov function. The procedure terminates once a valid Lyapunov function is found. Given that numerical root-finding algorithms do not provide guarantees, a formal verification can be applied to the final analytical Lyapunov function we found, which is a lot easier compared with verifying neural networks. Following the compositional neural certificate \cite{pmlr-v211-zhang23a}, we further propose a compositional neural Lyapunov function design to scale up our approach to networked systems. 

We evaluate our algorithm with various nonlinear systems including the pendulum system, path-following problem, Van Der Pol oscillator, rotating wheel pendulum, 3-bus power system, and problems from nonlinear systems textbook \cite{khalil2002nonlinear}. Our method demonstrated the efficiency and robustness in discovering valid analytical Lyapunov functions. 
Contributions can be summarized as follows:
\begin{itemize}
    \item We present the first algorithm that combines neural networks with symbolic regression to construct analytical Lyapunov functions directly, which significantly facilitates interpretability. 
    \item We propose an efficient counterexample generation paradigm by sampling around the roots of the symbolic Lyapunov functions, which circumvents expensive neural network verifiers.
    \item We validate our algorithm across various examples with neural networks with more than 100 neurons.
\end{itemize}
\section{Related Works}
\subsection{Learning-based Lyapunov function construction}
The area of learning-based Lyapunov function construction is experiencing significant growth. A prominent model in this field is the Neural Lyapunov Control \cite{chang2019neural}. This framework includes a neural-network-based Lyapunov function and a learnable linear controller, with stability verified by an SMT solver. This methodology was extended by \cite{zhou2022neural}, which introduced a neural network to model unknown dynamics and another for control. For switched affine systems, \cite{10543148} introduces a joint learning scheme for a common Lyapunov function and the controller. Additionally, \cite{dailyapunov,wu2023neural} consider discrete-time systems and neural network controllers, verifying stability via MIP solvers. One limitation of these methods is the need for linearization of dynamics or other relaxations for formal verification. \cite{yang2024lyapunovstable} deploys $\alpha,\beta$-CROWN for neural network verification due to its scalability, and extends the previous control scheme from state feedback to output feedback control. However, these formal verification methods face challenges with large neural networks: SMT solvers can handle up to 30 neurons, MIP solvers are limited to networks with up to 200 neurons \cite{10015199}, and \cite{yang2024lyapunovstable} has 16 neurons each layer for the Lyapunov neural network. 


\subsection{Symbolic regression}
Symbolic regression is a supervised machine-learning technique that constructs analytical models to represent datasets. It typically addresses a multi-objective optimization problem, balancing between minimizing prediction error and model complexity. Common practice involves employing genetic algorithms for solution finding, a brute-force approach whose computational complexity scales up exponentially with input dimension \cite{Schmidt2010}. To address this concern, \cite{NEURIPS2020_c9f2f917} proposes to combine neural networks with the genetic algorithm-based symbolic regression tool \emph{eureqa} \cite{doi:10.1126/science.1165893}. This method first learns a black-box model and then applies symbolic regression to find the mathematical formulation for scientific discovery. An alternative to generic algorithm-based techniques is “SINDy” (Sparse Identification of Nonlinear Dynamics) \cite{sindy}, which formulates expressions as linear combinations from a fixed dictionary of nonlinear functions. Additionally, \cite{petersen2021deep, NEURIPS2023_8ffb4e31} introduces deep generative models for symbolic regression and can achieve fast inference. 

Given the extensive array of available algorithms, as summarized in \cite{cranmer2023interpretable}, we selected the open-source \textit{PySR} \cite{cranmer2023interpretable} for symbolic regression in our algorithm. This choice is driven by \textit{PySR}'s user-friendly nature, its ability to optimize unknown constants, the provision of multiple solutions beyond mere accuracy, and the flexibility to customize operators and handle constraints. Other packages in the community like "SINDy" could also likely be used and achieve similar results. 


\section{Preliminary}
\subsection{Lyapunov Stability}
We consider the problem of constructing Lyapunov functions for autonomous nonlinear dynamical systems at an equilibrium point. Without loss of generality, we assume the equilibrium point is the origin. 
\begin{definition}[Dynamical systems]
\label{def:dynamical system}
An n-dimensional autonomous nonlinear dynamical system is defined as 
\begin{equation}\label{eq:nonlinear_dynamics}
    \frac{dx}{dt}=f(x)\,,x(0)=x_0,
\end{equation}
where $f:\mathcal{D}\to \mathbb{R}^n$ is a Lipschitz-continuous vector field, and $\mathcal{D}\subseteq \mathbb{R}^n$ is an open set containing the origin that defines the state space. Each $x(t)\in\mathcal{D}$ is a state vector. 
\end{definition}

\begin{definition}[Asymptotic stability]
    \label{def:asm_stb}
    The origin of system \eqref{eq:nonlinear_dynamics} is stabilized if for any $\epsilon>0$, there exists $\delta(\epsilon)>0$ such that $\lVert x(t)\rVert<\epsilon$, $\forall t\geq 0$
    if $\lVert x(0)\rVert<\delta$. The origin is asymptotically stable if it is stable and $\delta$ can be chosen such that $\lVert x(0)\rVert <\delta \implies \lim\limits_{t \to \infty}x(t)=0$ \cite{khalil2002nonlinear}.
\end{definition}
\begin{definition}[Lie derivative] The Lie derivative of a continuously differentiable scalar function $V:\mathcal{D}\to\mathbb{R}$ along the trajectory of \eqref{eq:nonlinear_dynamics} is 
\begin{equation}
    \label{eq:Lie}
    L_f V(x)=\sum_{i=1}^n\frac{\partial V}{\partial x_i}\frac{dx_i}{dt}=\sum_{i=1}^n\frac{\partial V}{\partial x_i}f_i(x).
\end{equation}
\end{definition}

\begin{proposition}[Lyapunov functions for asymptotic stability]\label{pro:lyap} Let $0$ be an equilibrium point for \eqref{eq:nonlinear_dynamics} and $\mathcal{D}\subseteq\mathbb{R}^n$ be a domain containing the origin. Let $V:\mathcal{D}\to\mathbb{R}$ be a continuously differentiable function such that
\begin{subequations}
    \begin{align}
        &V(0)=0 \text{ and } V(x)>0 \text{ in } D\backslash\{0\}\label{eq:Lyapunov},\\ 
        &L_f V(x)\leq 0 \text{ in } D.
    \end{align}
\end{subequations}
Then, $x=0$ is stable. Moreover, if 
\begin{equation}\label{eq:decrease}
    L_f V(x)< 0\text{ in } D\backslash \{0\},
\end{equation}
then the origin is asymptotically stable.
\end{proposition}
    
\subsection{Neural Lyapunov Network}
Following \cite{kolter2019learning}, a neural network parameterization for Lyapunov function is provided as follows,
\begin{equation}\label{eq:NN-Ly}
    V_\phi(x)=\sigma(g(F(x))-g(F(0)))+\epsilon \lVert x\rVert_2^2,
\end{equation}
where $F:\mathbb{R}^n\to \mathbb{R}^n$ is a continuously differentiable invertible function, $g(x)$ is an input-convex neural network (ICNN) \cite{amos2017input},  $\epsilon$ is a small positive constant, $\sigma(\cdot)$ is a smoothed ReLU function defined by
\[
  \sigma(x) = \left\{
     \begin{array}{@{}l@{\thinspace}l}
       0  & \text{  if } x\leq 0,\\
       \frac{x^2}{2d} & \text{  if } 0<x<d, \\
       x-\frac{d}{2} & \text{  otherwise}\\
     \end{array}
   \right.
\]
Following Proposition \ref{pro:lyap}, the Lyapunov function needs to be positive semi-definite and have no local optima except the origin (due to the negative Lie derivative condition \eqref{eq:decrease}). Convexity of the neural network can prevent the existence of local optima. To enlarge the search space to potential non-convex candidates, function $F$ is deployed as an input to ICNN $g(\cdot)$. The final Lyapunov function is shifted to guarantee positive semi-definiteness.

The neural Lyapunov function can be updated by minimizing the Lyapunov risk \cite{chang2019neural} defined as 
\begin{equation}\label{eq:loss}
    L(\phi)=\frac{1}{N}\sum_{i=1}^N\left (max(0,L_f V_\phi(x_i))\right ),
\end{equation}
where $x_i\in\mathcal{D}\,, \forall i\in[1,2,3,...,N]$ are counterexamples that violate the stability condition \eqref{eq:decrease}. By design of the neural Lyapunov function, $max(0,-V_\phi(x))=V_\phi^2(0)=0$, thus we only need to enforce the Lie derivative condition. 
\subsection{Symbolic Regression}
We use the symbolic regression package \textit{PySR} for deriving neural network's analytical expressions, which employs a genetic algorithm to stochastically assemble algebraic operators to fit a mathematical model to the given dataset. 

With a predefined set of operators, \textit{PySR} proposes a closed-form analytical expression at each complexity level, maximizing the probability of finding the most appropriate expression. For example, complexity 1 might yield $x^2$, complexity 2 results in $2x^2$, and complexity 3 produces $2x^2+y$, continuing up to the maximum depth. To align with the characteristics of the Lyapunov function, the operators used in our study include $+, -, *, /, \text{square}, \sin, \cos,$ and $1-\cos$, where $1-\cos$ is deployed as the non-negative counterparts of $\cos$. Our approach allows for the nesting of these operators and the optimization of constants. Expressions given by \textit{PySR} may look like $\sin(\sin(x-y))$ or $0.64*(\sin(x))^2*(1-\cos(y))+0.57z$, where $0.64$ and $0.57$ are examples of the constants optimized by \textit{PySR}.

\section{Proposed Method}
\begin{figure*}[h!]
    \centering
    \includegraphics[width=0.7\textwidth]{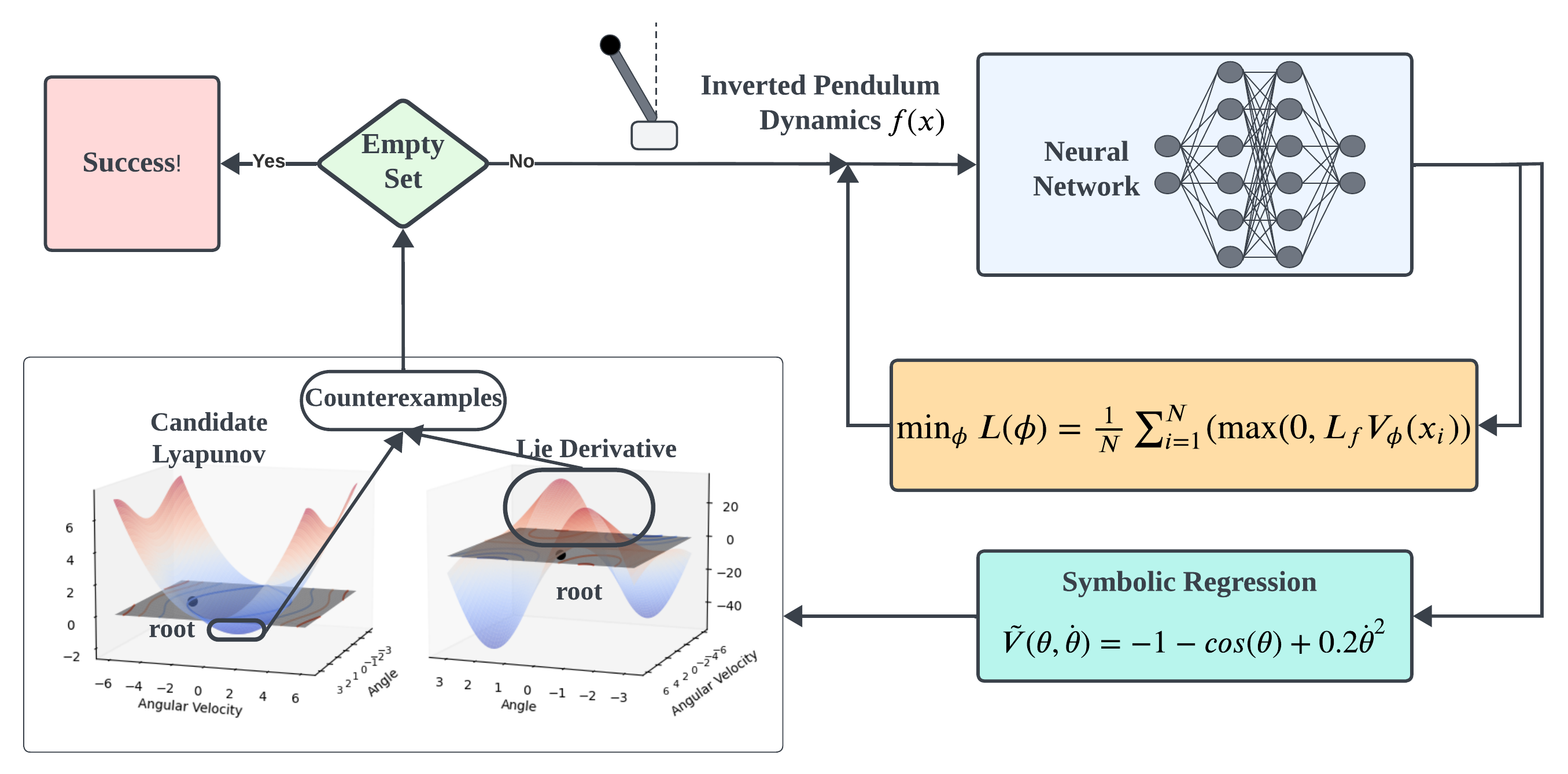}
    \caption{Diagram illustrating the proposed framework applied to an inverted pendulum. The analytical form (invalid intermediate result) enables root identification for both the candidate Lyapunov function and the Lie derivative. Counterexamples are sought around the roots. The process concludes when no counterexamples are detectable.}
    \label{fig:method}
\end{figure*}

In this section, we detail our proposed framework which can directly construct an analytical Lyapunov function, and thus can be rigorously verified locally and potentially generalized globally. We summarize the training paradigm in Algorithm \ref{alg:our} and present the framework in Figure \ref{fig:method}.

\subsection{Framework Overview}
We start by introducing the outlook for our framework. As demonstrated in Figure \ref{fig:method}, our framework contains two loops. The inner loop is the learner, which optimizes the neural network defined by \eqref{eq:NN-Ly} through minimizing the Lyapunov risk \eqref{eq:loss}. The outer loop is the falsifier, which deploys symbolic regression to retrieve an analytical formulation for the neural network, and then generates counterexamples for further optimization. The outer loop is triggered every \(N_{in}\) updates of the inner loop to avoid redundant regressions. This process, including the \(N_{in}\) updates and the symbolic regression, is referred to as one epoch.
For networked systems with high-dimensional states, we adopt the compositional neural Lyapunov function design in \cite{pmlr-v211-zhang23a} for scalability. In the remaining part, we detail the symbolic-regression-based falsifier and the compositional design.  
\subsection{Symbolic Regression and Falsification}
To facilitate the training of the neural Lyapunov network, we use the symbolic regression package \textit{PySR} to derive the neural network's analytical expressions. The process begins by uniformly sampling $N_p$ points within the state space, which are input into the neural network to generate corresponding outputs. These input-output pairs are then fed into \textit{PySR}, which provides $n$ symbolic formulations at different complexity levels modeling the neural network. Notably, those $N_p$ pairs are solely for fitting the symbolic regression model, not for neural network training. 


With the expressions proposed by \textit{PySR}, we evaluate the candidate Lyapunov functions with the Symbolic Python (SymPy) \cite{10.7717/peerj-cs.103} for direct computation of the closed-loop form derivative. Specifically, given $\Tilde{V}(x)$ provided by \textit{PySR}, we directly calculate $\frac{d\Tilde{V}(x)}{dx}$. This allows us to get the Lie derivative following \eqref{eq:Lie} with the known dynamics $f(x)$. Subsequently, we employ root-finding tools to identify roots for both the Lyapunov function and the Lie derivative. For root identification of both $\Tilde{V}(x)$ and $L_f\Tilde{V}(x)$, we apply SciPy's \textit{fsolve} function \cite{2020SciPy-NMeth} in its default setting. If nonzero roots are detected, the Lyapunov function is invalidated. Otherwise, we sample a random nonzero state $x\in\mathcal{D}$ to verify that $\Tilde{V}(x)>0$, which rules out the condition that the whole Lyapunov function $\Tilde{V}(x)$ is negative without roots in the considered region. A similar operation can validate $L_f\Tilde{V}(x)<0$. With a successful pass of the two simple verifications, $\Tilde{V}(x)$ is a numerically valid Lyapunov function for dynamics $f(x)$ in the interested domain. Notably, although our neural network design inherently ensures positive semi-definiteness, we do not restrict PySR to only produce PSD formulations. Non-PSD formulations can be valuable, sometimes requiring just constant optimization to get the valid Lyapunov function.

When nonzero roots are located, counterexamples can be generated from these roots. For instance, if a nonzero root $r$ satisfies $L_f \Tilde{V}(r)=0$, it is typically feasible to find $L_f \Tilde{V}(r+\epsilon)>0$ with some small enough $\epsilon$. Counterexamples are then created by progressively increasing $\epsilon$ until $L_f \Tilde{V}(r+\epsilon)<0$. While both \eqref{eq:Lyapunov} and \eqref{eq:decrease} are checked for the candidate Lyapunov functions, we only gather counterexamples violating \eqref{eq:decrease} for network learning. Additionally, we randomly sample points in the state space to identify other potential counterexamples. Recall that PySR proposes multiple different candidate functions, we generate counterexamples for each and combine them for the subsequent neural network training. The algorithm terminates once a numerically valid analytical Lyapunov function is found. We present a pseudocode of the proposed learning framework for Lyapunov function construction in Algorithm \ref{alg:our}.

\begin{remark}
   Without a perfect root finder, our method accelerates counterexample generation, leading to an analytical Lyapunov function. In experiments, we apply SMT to the found Lyapunov functions for final validation. 
\end{remark}
\begin{algorithm}[tb]
   \caption{CoNSAL (Combining Neural networks and Symbolic regression for Analytical Lyapunov function)}
   \label{alg:our}
\begin{algorithmic}[1]
   \STATE {\bfseries Input:} Dynamics $f(x)$, Training steps $N_{in}$, Maximum complexity $n$.
   \STATE Initialize the neural Lyapunov function $V_\phi(x)$,
   \STATE Randomly sample input points $x\in\mathcal{D}$,
   \REPEAT
   \FOR{$i=1$ {\bfseries to} $N_{in}$}
   \STATE $L_fV_\phi = \sum_{i=1}^n\frac{\partial V_\phi}{\partial x_i}f_i(x)$,    \COMMENT{Forward Pass}
   \STATE Compute Lyapunov risk $L(\phi)$,
   \STATE $\phi \leftarrow \phi-\eta \nabla_\phi L(\phi)$, \COMMENT{Back propagation}
   \ENDFOR
   \STATE $\{\Tilde{V}_1, \cdots \Tilde{V}_n\}\leftarrow \text{PySR}(V_\phi)$, \COMMENT{Symbolic Regression}
   \FOR{$j=1$ {\bfseries to} $n$} 
   \STATE Find roots for $\Tilde{V}_j$, \COMMENT{Falsification}
   \IF{Lyapunov conditions \eqref{eq:Lyapunov} and \eqref{eq:decrease} hold,}
   \STATE Lyapunov function $\Tilde{V}_j$ is valid, return.
   \ENDIF
   \STATE Generate counterexamples,
   \ENDFOR
   \STATE Concatenate $x$ with counterexamples generated by $\Tilde{V}_j$ for all $j\in[1,2,3...n]$,
   
   \UNTIL{Lyapunov function is valid.}
\end{algorithmic}
\end{algorithm}
\subsection{Compositional Neural Lyapunov Function}
Although this proposed algorithm can efficiently find the Lyapunov functions for relatively small systems, it's still an open challenge for neural Lyapunov functions to scale up to high-dimensional systems, e.g. large-scale network systems. In this paper, we consider networked systems involving $m$ subsystems $\mathcal{S}=[1,2,...,m]$ following \cite{pmlr-v211-zhang23a}, and the dynamics of each subsystems can be written as $\frac{dx_i}{dt}=f_i(x_i,x_{i_1},x_{i_2},...,u_i)=f_i(x_i,x_{\mathcal{N}_i},u_i)$, where $\mathcal{N}_i$ means the neighbor of subsystem $i$ including ${i_1,i_2,...}$ that are connected to subsystem $i$. With slight violation of notations, we use $x_i\in\mathbb{R}^{d_i}$ as the states of subsystem $i$ and $u_i$ as the corresponding actions. 

To address this issue, an intuitive way is to find compositional Lyapunov functions for the high-dimensional system instead of a single Lyapunov function, i.e. finding one Lyapunov function $V_i(x_i)$ for each subsystem $i$. Moreover, if the networked system has a similar symmetric structure for different subsystems, the local Lyapunov functions can share the same structure across different subsystems. Following \cite{pmlr-v211-zhang23a}, when the individual Lyapunov functions $V_i(x_i)$ satisfy an Input-to-State Stability (ISS)-style condition, they will certify the stability of the entire dynamical system. 

\begin{figure}[h]
    \centering
    \includegraphics[width=\linewidth]{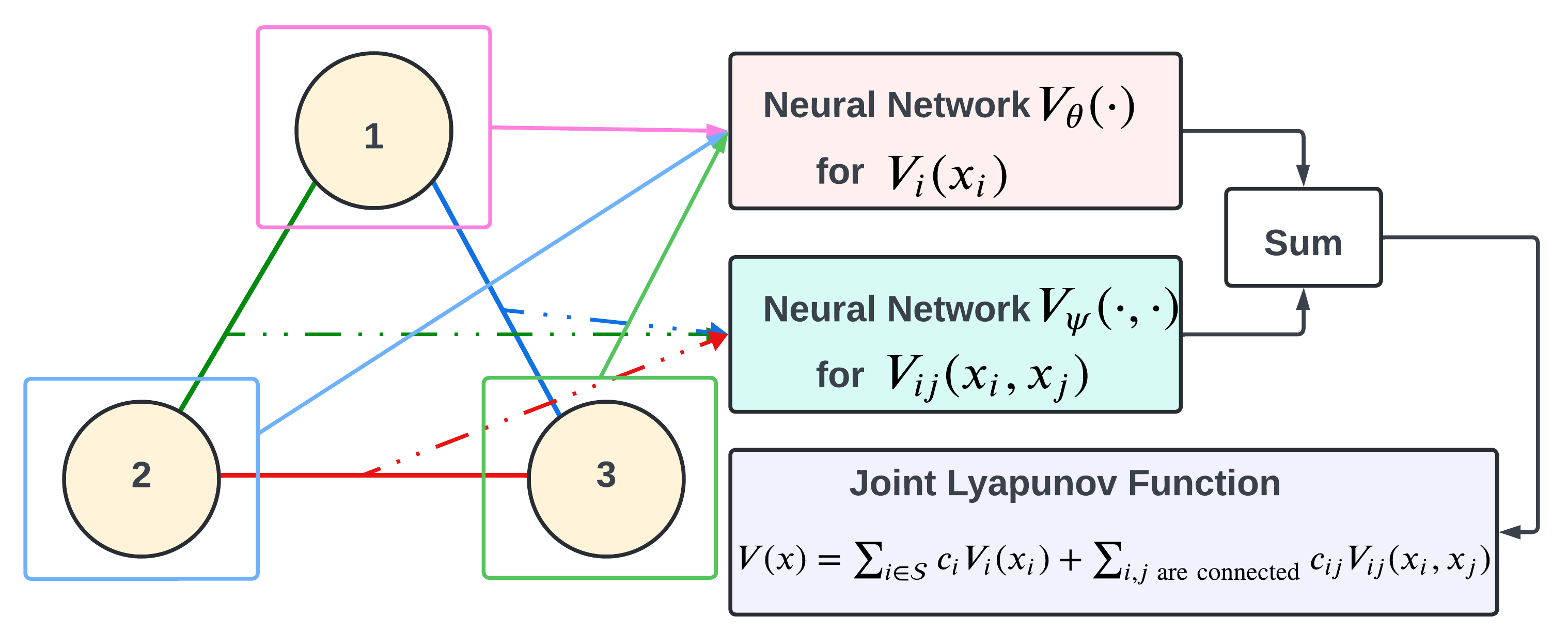}
    \caption{Diagram illustrating the proposed compositional structure for a three-node network. Each subsystem's states are fed to the neural network \( V_\theta(\cdot) \), and the connected subsystems' states are fed to \( V_{\psi}(\cdot, \cdot) \). These neural networks are shared across edges and subsystems, and the joint Lyapunov function for the entire system is obtained by summation.}
    \label{fig:compositional}
\end{figure}

Inspired by these results, we propose generalizing the compositional neural Lyapunov function design to certify stability. This approach ensures stability by considering the interrelationships between subsystems in addition to the individual Lyapunov functions for each subsystem. Instead of using one shared neural network to learn $V_i(x_i)$, we propose to include a supplementary neural network that models the interactions between different subsystems, i.e. $V_{ij}(x_i,x_j)$. $V_{ij}(x_i,x_j)$ can be regarded as the energy of the edge that connects subsystems $i,j$. For simplicity, we assume that each edge only connects two subsystems, each subsystem has the same state dimensions and shares symmetric structures, and the Lyapunov function for the whole system can be represented by $V(x)=\sum_{i\in\mathcal{S}} c_iV_i(x_i)+\sum_{i\in\mathcal{S}}\sum_{j\in\mathcal{N}_i} c_{ij}V_{ij}(x_i,x_j)$, where $c_i,c_{ij}$ are some constants. These assumptions are satisfied by the power system dynamics \cite{cui2023structured} and the Platoon system \cite{pmlr-v211-zhang23a}. In this case, we achieve a significant reduction of input dimensions, which simplifies the training of neural networks and the following symbolic regression. Specifically, we model $V_\theta(x_i)$ with parameter $\theta$ for all $V_i(x_i)$ and $V_\psi(x_i,x_j)$  with parameter $\psi$ for $V_{ij}(x_i,x_j)$, $c_i,c_{ij}$ are also set as learnable parameters. To reduce the number of neural networks, parameters of neural network that models $V_i(\cdot)$ are shared across m subsystems and so do those modeling $V_{ij}(\cdot,\cdot)$. The neural network structure is presented in Figure \ref{fig:compositional}

\begin{figure*}
            \centering
            \begin{minipage}[t]{0.3\textwidth}
                \centering
                \includegraphics[width=\linewidth]{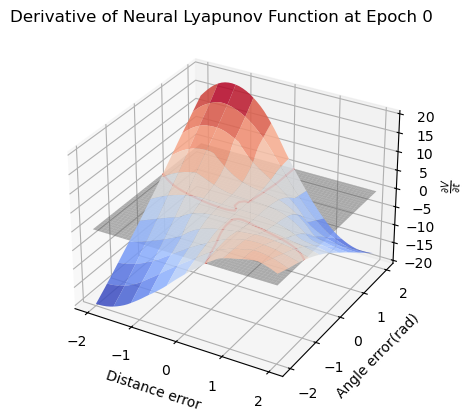}
            \end{minipage}
            \hfill
            \begin{minipage}[t]{0.3\textwidth}
                \centering
                \includegraphics[width=\linewidth]{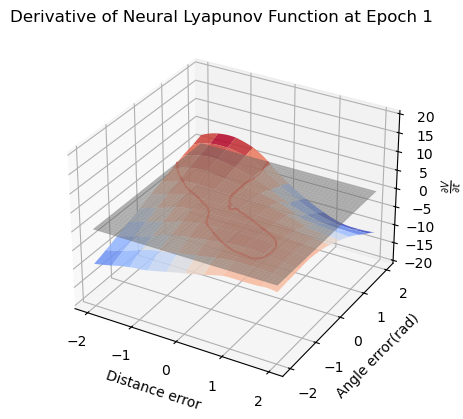}
            \end{minipage}
            \hfill
            \begin{minipage}[t]{0.3\textwidth}
                \centering
                \includegraphics[width=\linewidth]{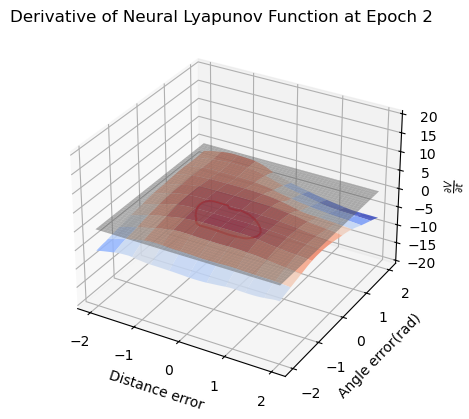}
            \end{minipage}
            \caption {Lie Derivative of the Neural Lyapunov Function under linear path following dynamics, updated with the proposed algorithm. Showing the result of the first two epochs.}
            \label{fig: epoch 100}
    \label{fig: Neural Lyapunov Function Evolution}
    \end{figure*}

We train and validate the compositional Lyapunov function with a similar scheme as the low-dimensional systems. During training, we optimize the Lyapunov risk \eqref{eq:loss} with the joint Lyapunov function $V(x)$ to jointly update the neural Lyapunov functions and the coefficients $c_i,c_{ij}$. After several updating steps, symbolic regression is triggered to reveal the analytical formulation of $V_i(x_i)$ and $V_{ij}(x_i,x_j)$, which are then combined with the coefficients to get $V(x)$. The joint Lyapunov function is then validated by the same process. If it is not valid, we generate counterexamples around the roots and continue the iterative training.

\section{Experiment}
In this section, we validate the proposed algorithm with various nonlinear dynamics by finding their Lyapunov functions, where the autonomous systems (without additional control) are stable at the origin. We test the proposed algorithm using the neural Lyapunov network \eqref{eq:NN-Ly} on 2-D path following, inverted pendulum, Van Der Pol oscillator, 3-D trig dynamics, 4-D rotating wheel pendulum, and 6-D nonlinear dynamics. The compositional neural Lyapunov function is applied to a 3-bus power system. To ease the requirement for symbolic regression to find the exact constants in the Lyapunov functions and accommodate the error of numerical root-finding tools, we allow a violation of the stability conditions less than $10^{-4}$ in value and the Lyapunov function is only verified at a given local region numerically. CoNSAL aims to find an analytically correct Lyapunov function with tolerable errors. Global stability can be assessed through further inspection. In this work, multiple initial guesses are applied to the root finder to facilitate root finding. 

We first demonstrate the evolution of the Neural Lyapunov function, and compare how the proposed framework optimizes the neural Lyapunov function with the SMT-based scheme \cite{chang2019neural} in the 2D path following environment, where our approach can achieve better performance with respect to the SMT solver. 
The runtime information and the found analytical Lyapunov functions are summarized in Table \ref{tab:performance summary}. Details about dynamics are available in Appendix \ref{apdx:dynamics}. Neural network configurations and base operator selection are available in Appendix \ref{apdx:neural}. 
\subsection{Evolution of Neural Lyapunov Function}
Linear path following is a classic nonlinear control task, which controls a wheeled robot to follow a linear path with dynamics \eqref{eq:linear} given by \cite{path-following}:

\begin{subequations}\label{eq:linear}
    \begin{align}
    \Dot{x}_1 &= v\cdot sin(x_2), \\
    \Dot{x}_2 &= - x_2- c \cdot v \cdot \frac{sin(x_2)}{x_2} \cdot x_1,
\end{align}
\end{subequations}
where $x_1$, $x_2$ are the distance and the and angel difference between the robot and the reference line, $v\in \mathbb{R^+}$ is a constant velocity and $c \in \mathbb{R^+}$ is a positive real constant. Specifically, we set $c=2, v= 6 \si{\meter\per\second}$. We consider state space $\mathcal{D} = \{ (x_1, x_2) \in \mathbb{R}^2 \bm{\mid} |x_1| \leq 2, |x_2| \leq \pi \}$. This system is stable to the origin and has a known Lyapunov function $V(x_1, x_2) = x_{1}^2 + \frac{x_{2}^2}{c}$, which can guarantee global asymptotic stability. With this system, we illustrate the evolution of the neural Lyapunov function in Figure \ref{fig: epoch 100}. As shown in the figure, the landscape of the Lie derivative evolves rapidly to satisfy the stability condition, and the violation of the negative Lie derivative decreases efficiently, demonstrating the effectiveness of our proposed framework. 

We further compare our framework with the SMT-based neural Lyapunov function \cite{chang2019neural}, where the SMT solver is deployed to generate counterexamples for the neural network and validate the stability conditions. We use the same two-layer neural networks with 128 hidden units for both methods. Given the relatively large neural network, the SMT-based approach can only achieve 30\% successful rate for converging, while ours achieved 80 \% successful rate for finding the valid symbolic Lyapunov function. Following Table \ref{table: comparison summary}, due to the complexity of the neural network, the SMT solver consumes much more time compared with symbolic-regression-based verification.

\begin{figure*}
    \centering
    \begin{minipage}[t]{0.3\textwidth}
        \centering
        \includegraphics[width=\linewidth]{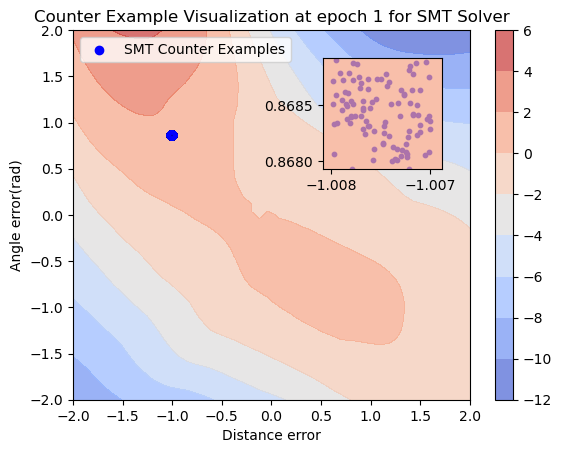}
    \end{minipage}
    \hfill
    \begin{minipage}[t]{0.3\textwidth}
        \centering
        \includegraphics[width=\linewidth]{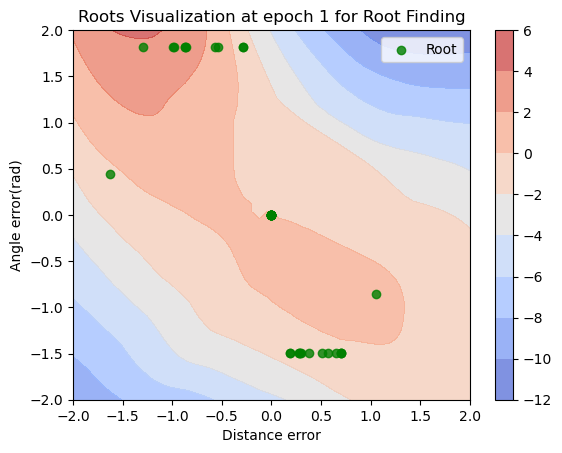}
    \end{minipage}
    \hfill
    \begin{minipage}[t]{0.3\textwidth}
        \centering
        \includegraphics[width=\linewidth]{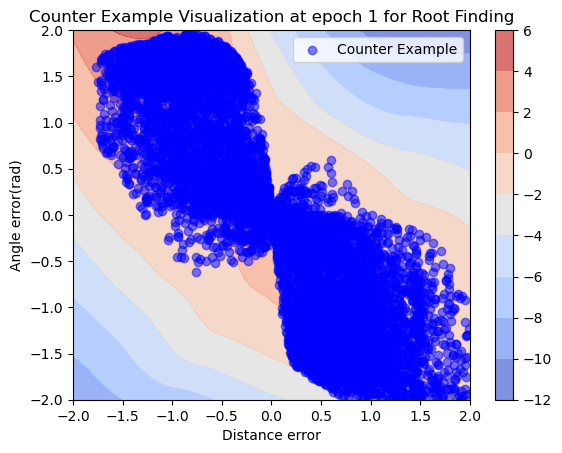}
    \end{minipage}
    \caption {SMT Counter Examples, Roots, and Root finding Counter Examples Visualization using the same checkpoint at epoch 1 for Lie derivative of the neural Lyapunov functions under linear path-following dynamics. The same number of counterexamples are generated. The zoomed-in region shows the counterexamples from SMT solvers. }
    \label{fig: ce smt}
\end{figure*}

The counterexamples generated by both algorithms are presented in Figure \ref{fig: ce smt}. Following this result, the SMT solver can identify a small region such that samples from the region will violate the stability constraint. Thus counterexamples are generated by randomly sampling in this region. However, this method provides counterexamples focusing on this small region and can potentially lead to overfitting. With PySR, multiple symbolic expressions are generated at different complexity levels. The roots of these formulations can be efficiently found numerically, which are illustrated in the middle plot of Figure \ref{fig: ce smt}. With acceptable approximating errors, the roots outline the region of violation and the counterexamples can be efficiently generated near the roots. In this case, our approach finds counterexamples covering most of the violation areas in the state space and facilitates the successful discovery of a valid Lyapunov function. 

\begin{table}[h]
\centering
\caption{Runtime and Performance Comparison}
\label{table: comparison summary}
\begin{threeparttable}
\begin{tabular}{|p{3cm}|p{1.8cm}|p{1.8cm}|}
   \hline
    \multicolumn{3}{|c|}{\textbf{Verification Methods Comparison}} \\
    \hline
    \diagbox[width=3.4cm]{\textbf{Metric}}{\textbf{Method}} & Ours & SMT Solver\\
    \hline
    Verification Time & 5.81s & 92.08s \\
    \hline
    Convergence Time & 1401s & 5051s \\
    \hline
    Convergence Epoch & 45 & 58 \\
    \hline
    Success Rate & 80\% & 30\% \\
    \hline
\end{tabular}
\begin{tablenotes}
    \tiny
    \item \textit{Note: Verification time measures the average time consumption for each single-step verification. Only success trials' convergence time and convergence epoch are included in this table. The success rate is calculated out of 10 different trials. An experiment that found a valid Lyapunov function within 100 epochs is considered a successful trial.}
\end{tablenotes}
\end{threeparttable}
\end{table}

\begin{figure}[h]
    \includegraphics[width=0.35\textwidth]{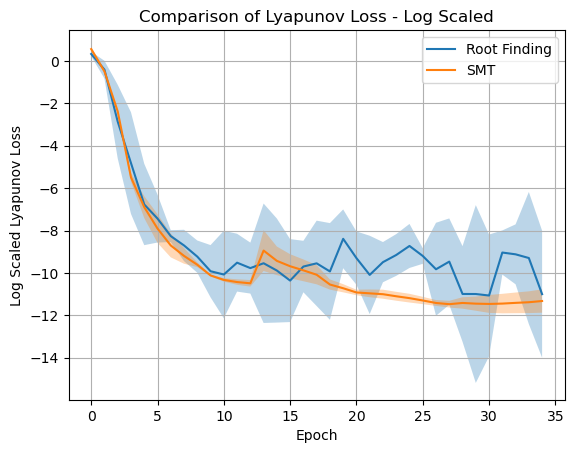}
    \caption{Log-scaled Loss Convergence Comparison using loss \eqref{eq:loss}. Only success trials are considered.}
    \label{fig:loss}
\end{figure}

We conclude this comparison with the loss curve in Figure \ref{fig:loss}. Given that our approach covers a larger violation space than the SMT-based method, our algorithm can induce a greater loss calculated by \eqref{eq:loss} and a larger variance. However, our method can still achieve comparable loss convergence compared with the SMT-based approach. Empirically we also find that, if the SMT-based neural Lyapunov function fails to converge, PySR cannot find the analytical Lyapunov function from the SMT-trained neural network even if it has a low Lyapunov risk.

\begin{table*}[t]
\centering 
\begin{tabular}{@{}llllll@{}}
\toprule
                      \textbf{Dynamics}& \textbf{Time}& \textbf{Epoch} & \textbf{Found Lyapunov Functions$^a$} & \textbf{Stability}$^\ddagger$
                      &\textbf{Succ \% $^\ddagger$}\\ \midrule
\multicolumn{1}{l|}{2-D Path Following (Linear)} &  1401s & 45  & $V=x_1^2 + \frac{x_2^2}{c}$ & g.a.s. & 80\% \\
\multicolumn{1}{l|}{2-D Inverted Pendulum } & 1356s & 37 & $V = 1 - cos(x_1) + 0.254x_2^2$ & l.a.s. & 60\% \\
\multicolumn{1}{l|}{2-D Van Der Pol Oscillator} &   230s  & 7 & $V=x_1^2 + x_{2}^2$ & l.a.s. & 100\% \\ 
\multicolumn{1}{l|}{3-D Trig Dynamics}  &   1216s  & 41 & $V = 2sin(x_1)^2 + x_2^2 + 2sin(x_3)^2$ & l.a.s. & 100\%\\ 
\multicolumn{1}{l|}{4-D Rotating Wheel Pendulum}  & 1556s & 34 & 
        $V = E^2+x_{3}^2+x_{4}^2$ $^{*}$ & g.a.s. & 80\%\\ 
\multicolumn{1}{l|}{6-D Nonlinear Dynamics}  & 916s & 13 & $\begin{array}[t]{l}
V = 4x_1^2 - 1.24x_2^2 + \left(\sum\limits_{i=1}^{5}x_i\right)^2 \\
\quad + 2.996 \left(0.578\sum\limits_{i=2}^{6}x_i^2 + 1\right)^2
\end{array}$ & l.a.s. & 100\%\\ 
\multicolumn{1}{l|}{6-D 3-bus Power System}  & 271s & 3 & $\begin{array}[t]{l}
V = 0.516  \left(\sum\limits_{i=1}^{3} \omega_{i}^2 \right) -\\ 0.5\sum\limits_{i=1}^{3} \sum\limits_{\substack{j=1 \\ i\neq j}}^{3} \left(cos(\delta_i - \delta_j) - 1 \right)
\end{array}$ & l.a.s. & 60\% \\                   
                      \bottomrule
                      \multicolumn{6}{l}{\tiny $a$. All Found Lyapunov functions passed SMT solver's verification for Lyapunov conditions, with tolerable error $\epsilon = 10^{-4}$.} \\
                      \multicolumn{6}{l}{\tiny $*$. $E$ is the energy function of the corresponding dynamics, which is defined at \ref{sub:Rotating Wheel Pendulum}.} \\
                      \multicolumn{6}{l}{\tiny  $\ddagger$. In this column, `g.a.s' represents globally asymptotically stable, and `l.a.s.' represents locally asymptotically stable.}\\
                      \multicolumn{6}{l}{\tiny $\ddagger$. 'Succ \%' denotes the successful rate of finding a valid Lyapunov function out of 5 random seeds. } \\
                      \multicolumn{6}{l}{\tiny Only success trials' convergence time and convergence epoch are included in this table. }

\end{tabular}
\caption{Training time and the found analytical Lyapunov functions for the proposed algorithm with different dynamics.}
\label{tab:performance summary}
\end{table*}
\vspace{-0.5cm}
\subsection{Runtime and Found Analytical Lyapunov Functions}
We summarize the runtime and the found Lyapunov functions in table \ref{tab:performance summary}. In general, the proposed framework can achieve more than $60\%$ success rate for all the tested dynamics and find at least a local Lyapunov function within $50$ symbolic regression calls, where a local Lyapunov function means the Lyapunov function is valid in the tested local region. Global stability is further extended by human experts. Notably, our algorithm demonstrates fast convergence for the tested 4-D and 6-D systems. We also find that our method can occasionally discover new Lyapunov functions valid locally (detailed in \ref{sub:vanderpol}). Here, `new' means that the Lyapunov function is not included in the existing literature.

However, it gets the lowest success rate with the 2-D inverted pendulum system and the 6-D 3-bus power system. The challenge comes from optimizing the constants in the Lyapunov function based on given physical constants in the dynamics. Considering the inverted pendulum system, its local Lyapunov function can be represented as 
$V(x_1, x_2)=(1 - cos(x_1)) + \frac{l}{2g} x_2 ^ 2$, where $x_1$ is the angular position from the inverted position, $x_2$ is the angular velocity, and $g, m, l, b$ are acceleration of gravity, mass of inverted object, the length of string, and the coefficient of friction respectively. Setting $l=5\si{\meter}$, $g=9.81m/ s^2$, it is challenging for the symbolic regression to find the correct coefficients. In failed cases, our framework often finds the correct structure but incorrect coefficients, e.g. $V=1-cos(x_1)+0.26x_2^2$, which is not valid. 

In cases where prior dynamics knowledge is available, we incorporate this knowledge into the Lyapunov function design to ease this task. Specifically, for the 4-D rotating wheel pendulum, we use the energy function $E$ as part of the input to the neural Lyapunov function, along with all state variables. As a result, a Lyapunov based on the energy function is revealed by our method. Details of all tested dynamics and considered local regions are provided in Appendix \ref{apdx:dynamics}.

\section{Conclusion}
 In this paper, we propose CoNSAL, a framework to construct analytical Lyapunov functions for stable nonlinear dynamics. Our approach integrates a neural network learner with a symbolic regression solver, which extracts an analytical formula from the neural Lyapunov function. This symbolic Lyapunov function is further validated with numerical root-finding approaches. When the Lyapunov function is invalid, we efficiently sample counterexamples around the roots to facilitate the training process. The resulting valid Lyapunov function can potentially demonstrate global stability or aid experts in designing controllers. Compared to existing work, CoNSAL provides an analytical Lyapunov function instead of neural ones, which significantly improves the interpretability and generalizability of the final results. Moreover, it bypasses the need for neural network verification, enabling the use of larger neural networks and allowing for scalability to more complex systems. The efficiency of the proposed framework is validated in multiple nonlinear dynamical systems by successfully finding a valid local Lyapunov function for these test cases.

Our results open up several exciting future directions: (1) Efficiently incorporating known constants from the dynamics into symbolic regression to reveal analytical formulations without scaling up the input size;
 (2) Exploring neural network-based symbolic regression methods for fast inference and end-to-end training in Lyapunov function discovery without the requirement for base operators; (3) Investigating formal verification for analytical formulations beyond numerical root-finding; (4) Scaling up to larger dynamical systems with more than 10 dimensions; (5) Integrating a nonlinear controller in our framework and optimizing it alongside the Lyapunov function in an end-to-end manner, ensuring stability guarantees; (6) Maximizing the region of attraction. These directions can enhance the efficiency and generalizability of CoNSAL, making it applicable to the vast cases where analytical Lyapunov functions are valuable for validation, analysis, or controller design.
\section*{Acknowledgements}
This work was supported under grants NSF ECCS-2200692 and Jacobs School Early Career Faculty Development Award 2023.
Jie Feng is supported by the UC-National Laboratory In-Residence Graduate Fellowship L24GF7923. Haohan Zou is funded by UC San Diego Halıcıoğlu Data Science Institute (HDSI) scholarship. Yuanyuan Shi is partially supported by the Hellman Fellowship.

\pagebreak
\nocite{langley00}

\bibliography{example_paper}

\begin{thebibliography}{36}
\providecommand{\natexlab}[1]{#1}
\providecommand{\url}[1]{\texttt{#1}}
\expandafter\ifx\csname urlstyle\endcsname\relax
  \providecommand{\doi}[1]{doi: #1}\else
  \providecommand{\doi}{doi: \begingroup \urlstyle{rm}\Url}\fi

\bibitem[Ahmadi \& Majumdar(2016)Ahmadi and Majumdar]{sos}
Ahmadi, A.~A. and Majumdar, A.
\newblock Some applications of polynomial optimization in operations research and real-time decision making.
\newblock \emph{Optimization Letters}, pp.\  709--729, 2016.

\bibitem[Amodei et~al.(2016)Amodei, Olah, Steinhardt, Christiano, Schulman, and Mané]{amodei2016concrete}
Amodei, D., Olah, C., Steinhardt, J., Christiano, P., Schulman, J., and Mané, D.
\newblock Concrete problems in ai safety, 2016.

\bibitem[Amos et~al.(2017)Amos, Xu, and Kolter]{amos2017input}
Amos, B., Xu, L., and Kolter, J.~Z.
\newblock Input convex neural networks.
\newblock In \emph{International Conference on Machine Learning}, pp.\  146--155. PMLR, 2017.

\bibitem[Angelis et~al.(2023)Angelis, Sofos, and Karakasidis]{angelis2023artificial}
Angelis, D., Sofos, F., and Karakasidis, T.~E.
\newblock Artificial intelligence in physical sciences: Symbolic regression trends and perspectives.
\newblock \emph{Archives of Computational Methods in Engineering}, pp.\  1--21, 2023.

\bibitem[Brunton et~al.(2016)Brunton, Proctor, and Kutz]{sindy}
Brunton, S.~L., Proctor, J.~L., and Kutz, J.~N.
\newblock Discovering governing equations from data by sparse identification of nonlinear dynamical systems.
\newblock \emph{Proceedings of the National Academy of Sciences}, 113\penalty0 (15):\penalty0 3932--3937, 2016.

\bibitem[Chang et~al.(2019)Chang, Roohi, and Gao]{chang2019neural}
Chang, Y.-C., Roohi, N., and Gao, S.
\newblock Neural lyapunov control.
\newblock \emph{Advances in neural information processing systems}, 32, 2019.

\bibitem[Cranmer(2023)]{cranmer2023interpretable}
Cranmer, M.
\newblock Interpretable machine learning for science with pysr and symbolicregression.jl, 2023.

\bibitem[Cranmer et~al.(2020)Cranmer, Sanchez~Gonzalez, Battaglia, Xu, Cranmer, Spergel, and Ho]{NEURIPS2020_c9f2f917}
Cranmer, M., Sanchez~Gonzalez, A., Battaglia, P., Xu, R., Cranmer, K., Spergel, D., and Ho, S.
\newblock Discovering symbolic models from deep learning with inductive biases.
\newblock In Larochelle, H., Ranzato, M., Hadsell, R., Balcan, M., and Lin, H. (eds.), \emph{Advances in Neural Information Processing Systems}, volume~33, pp.\  17429--17442. Curran Associates, Inc., 2020.

\bibitem[Cui et~al.(2023{\natexlab{a}})Cui, Jiang, and Zhang]{cuipower}
Cui, W., Jiang, Y., and Zhang, B.
\newblock Reinforcement learning for optimal primary frequency control: A lyapunov approach.
\newblock \emph{IEEE Transactions on Power Systems}, 38\penalty0 (2):\penalty0 1676--1688, 2023{\natexlab{a}}.
\newblock \doi{10.1109/TPWRS.2022.3176525}.

\bibitem[Cui et~al.(2023{\natexlab{b}})Cui, Jiang, Zhang, and Shi]{cui2023structured}
Cui, W., Jiang, Y., Zhang, B., and Shi, Y.
\newblock Structured neural-{PI} control with end-to-end stability and output tracking guarantees.
\newblock In \emph{Thirty-seventh Conference on Neural Information Processing Systems}, 2023{\natexlab{b}}.

\bibitem[Dai \& Permenter(2023)Dai and Permenter]{dai2023convex}
Dai, H. and Permenter, F.
\newblock Convex synthesis and verification of control-lyapunov and barrier functions with input constraints.
\newblock In \emph{2023 American Control Conference (ACC)}, pp.\  4116--4123. IEEE, 2023.

\bibitem[Dai et~al.(2021)Dai, Landry, Yang, Pavone, and Tedrake]{dailyapunov}
Dai, H., Landry, B., Yang, L., Pavone, M., and Tedrake, R.
\newblock Lyapunov-stable neural-network control.
\newblock In \emph{Proceedings of Robotics: Science and Systems}, 2021.

\bibitem[Dawson et~al.(2023)Dawson, Gao, and Fan]{10015199}
Dawson, C., Gao, S., and Fan, C.
\newblock Safe control with learned certificates: A survey of neural lyapunov, barrier, and contraction methods for robotics and control.
\newblock \emph{IEEE Transactions on Robotics}, 39\penalty0 (3):\penalty0 1749--1767, 2023.
\newblock \doi{10.1109/TRO.2022.3232542}.

\bibitem[Edwards et~al.(2024)Edwards, Peruffo, and Abate]{edwards2024fossil}
Edwards, A., Peruffo, A., and Abate, A.
\newblock Fossil 2.0: Formal certificate synthesis for the verification and control of dynamical models, 2024.

\bibitem[Fantoni \& Lozano(2002)Fantoni and Lozano]{wheel-pendulum}
Fantoni, I. and Lozano, R.
\newblock The reaction wheel pendulum.
\newblock 01 2002.
\newblock \doi{10.1007/978-1-4471-0177-2_7}.

\bibitem[Feng et~al.(2023{\natexlab{a}})Feng, Cui, Cortés, and Shi]{10163934}
Feng, J., Cui, W., Cortés, J., and Shi, Y.
\newblock Bridging transient and steady-state performance in voltage control: A reinforcement learning approach with safe gradient flow.
\newblock \emph{IEEE Control Systems Letters}, 7:\penalty0 2845--2850, 2023{\natexlab{a}}.
\newblock \doi{10.1109/LCSYS.2023.3289435}.

\bibitem[Feng et~al.(2023{\natexlab{b}})Feng, Shi, Qu, Low, Anandkumar, and Wierman]{10336939}
Feng, J., Shi, Y., Qu, G., Low, S.~H., Anandkumar, A., and Wierman, A.
\newblock Stability constrained reinforcement learning for decentralized real-time voltage control.
\newblock \emph{IEEE Transactions on Control of Network Systems}, pp.\  1--12, 2023{\natexlab{b}}.
\newblock \doi{10.1109/TCNS.2023.3338240}.

\bibitem[Feng et~al.(2024)Feng, Muralidharan, Henriquez-Auba, Hidalgo-Gonzalez, and Shi]{10543148}
Feng, J., Muralidharan, M., Henriquez-Auba, R., Hidalgo-Gonzalez, P., and Shi, Y.
\newblock Stability-constrained learning for frequency regulation in power grids with variable inertia.
\newblock \emph{IEEE Control Systems Letters}, 8:\penalty0 994--999, 2024.

\bibitem[Giesl \& Hafstein(2015)Giesl and Hafstein]{Peter}
Giesl, P. and Hafstein, S.
\newblock Review on computational methods for lyapunov functions.
\newblock \emph{Discrete and Continuous Dynamical Systems - B}, 20\penalty0 (8):\penalty0 2291--2331, 2015.
\newblock ISSN 1531-3492.
\newblock \doi{10.3934/dcdsb.2015.20.2291}.

\bibitem[Grüne(2019)]{6ddynamics}
Grüne, L.
\newblock Computing lyapunov functions using deep neural networks.
\newblock \emph{Journal of Computational Dynamics}, 8, 01 2019.
\newblock \doi{10.3934/jcd.2021006}.

\bibitem[Khalil(2002)]{khalil2002nonlinear}
Khalil, H.~K.
\newblock \emph{Nonlinear systems}.
\newblock Prentice Hall, 3nd edition, 2002.

\bibitem[Kolter \& Manek(2019)Kolter and Manek]{kolter2019learning}
Kolter, J.~Z. and Manek, G.
\newblock Learning stable deep dynamics models.
\newblock \emph{Advances in neural information processing systems}, 32, 2019.

\bibitem[Langley(2000)]{langley00}
Langley, P.
\newblock Crafting papers on machine learning.
\newblock In Langley, P. (ed.), \emph{Proceedings of the 17th International Conference on Machine Learning (ICML 2000)}, pp.\  1207--1216, Stanford, CA, 2000. Morgan Kaufmann.

\bibitem[McGough et~al.(2010)McGough, Christianson, and Hoover]{mcgough2010symbolic}
McGough, J.~S., Christianson, A.~W., and Hoover, R.~C.
\newblock Symbolic computation of lyapunov functions using evolutionary algorithms.
\newblock In \emph{Proceedings of the 12th IASTED international conference}, volume~15, pp.\  508--515, 2010.

\bibitem[Meurer et~al.(2017)Meurer, Smith, Paprocki, \v{C}ert\'{i}k, Kirpichev, Rocklin, Kumar, Ivanov, Moore, Singh, Rathnayake, Vig, Granger, Muller, Bonazzi, Gupta, Vats, Johansson, Pedregosa, Curry, Terrel, Rou\v{c}ka, Saboo, Fernando, Kulal, Cimrman, and Scopatz]{10.7717/peerj-cs.103}
Meurer, A., Smith, C.~P., Paprocki, M., \v{C}ert\'{i}k, O., Kirpichev, S.~B., Rocklin, M., Kumar, A., Ivanov, S., Moore, J.~K., Singh, S., Rathnayake, T., Vig, S., Granger, B.~E., Muller, R.~P., Bonazzi, F., Gupta, H., Vats, S., Johansson, F., Pedregosa, F., Curry, M.~J., Terrel, A.~R., Rou\v{c}ka, v., Saboo, A., Fernando, I., Kulal, S., Cimrman, R., and Scopatz, A.
\newblock Sympy: symbolic computing in python.
\newblock \emph{PeerJ Computer Science}, 3:\penalty0 e103, January 2017.
\newblock ISSN 2376-5992.
\newblock \doi{10.7717/peerj-cs.103}.

\bibitem[Petersen et~al.(2021)Petersen, Larma, Mundhenk, Santiago, Kim, and Kim]{petersen2021deep}
Petersen, B.~K., Larma, M.~L., Mundhenk, T.~N., Santiago, C.~P., Kim, S.~K., and Kim, J.~T.
\newblock Deep symbolic regression: Recovering mathematical expressions from data via risk-seeking policy gradients.
\newblock In \emph{International Conference on Learning Representations}, 2021.

\bibitem[Samson(1992)]{path-following}
Samson, C.
\newblock Path following and time-varying feedback stabilization of a wheeled mobile robot.
\newblock \emph{Second International Conference on Automation, Robotics and Computer Vision}, 3, 01 1992.

\bibitem[Schmidt \& Lipson(2009)Schmidt and Lipson]{doi:10.1126/science.1165893}
Schmidt, M. and Lipson, H.
\newblock Distilling free-form natural laws from experimental data.
\newblock \emph{Science}, 324\penalty0 (5923):\penalty0 81--85, 2009.
\newblock \doi{10.1126/science.1165893}.

\bibitem[Schmidt \& Lipson(2010)Schmidt and Lipson]{Schmidt2010}
Schmidt, M. and Lipson, H.
\newblock \emph{Symbolic Regression of Implicit Equations}, pp.\  73--85.
\newblock Springer US, Boston, MA, 2010.
\newblock ISBN 978-1-4419-1626-6.
\newblock \doi{10.1007/978-1-4419-1626-6_5}.

\bibitem[Shojaee et~al.(2023)Shojaee, Meidani, Barati~Farimani, and Reddy]{NEURIPS2023_8ffb4e31}
Shojaee, P., Meidani, K., Barati~Farimani, A., and Reddy, C.
\newblock Transformer-based planning for symbolic regression.
\newblock In \emph{Advances in Neural Information Processing Systems}, volume~36, pp.\  45907--45919. Curran Associates, Inc., 2023.

\bibitem[Virtanen et~al.(2020)Virtanen, Gommers, Oliphant, Haberland, Reddy, Cournapeau, Burovski, Peterson, Weckesser, Bright, {van der Walt}, Brett, Wilson, Millman, Mayorov, Nelson, Jones, Kern, Larson, Carey, Polat, Feng, Moore, {VanderPlas}, Laxalde, Perktold, Cimrman, Henriksen, Quintero, Harris, Archibald, Ribeiro, Pedregosa, {van Mulbregt}, and {SciPy 1.0 Contributors}]{2020SciPy-NMeth}
Virtanen, P., Gommers, R., Oliphant, T.~E., Haberland, M., Reddy, T., Cournapeau, D., Burovski, E., Peterson, P., Weckesser, W., Bright, J., {van der Walt}, S.~J., Brett, M., Wilson, J., Millman, K.~J., Mayorov, N., Nelson, A. R.~J., Jones, E., Kern, R., Larson, E., Carey, C.~J., Polat, {\.I}., Feng, Y., Moore, E.~W., {VanderPlas}, J., Laxalde, D., Perktold, J., Cimrman, R., Henriksen, I., Quintero, E.~A., Harris, C.~R., Archibald, A.~M., Ribeiro, A.~H., Pedregosa, F., {van Mulbregt}, P., and {SciPy 1.0 Contributors}.
\newblock {{SciPy} 1.0: Fundamental Algorithms for Scientific Computing in Python}.
\newblock \emph{Nature Methods}, 17:\penalty0 261--272, 2020.
\newblock \doi{10.1038/s41592-019-0686-2}.

\bibitem[Wang et~al.(2024)Wang, Andersson, and Tron]{wang2024lyapunov}
Wang, Z., Andersson, S.~B., and Tron, R.
\newblock Lyapunov neural network with region of attraction search, 2024.

\bibitem[Wu et~al.(2023)Wu, Clark, Kantaros, and Vorobeychik]{wu2023neural}
Wu, J., Clark, A., Kantaros, Y., and Vorobeychik, Y.
\newblock Neural lyapunov control for discrete-time systems.
\newblock In \emph{Thirty-seventh Conference on Neural Information Processing Systems}, 2023.

\bibitem[Yang et~al.(2024)Yang, Dai, Shi, Hsieh, Tedrake, and Zhang]{yang2024lyapunovstable}
Yang, L., Dai, H., Shi, Z., Hsieh, C.-J., Tedrake, R., and Zhang, H.
\newblock Lyapunov-stable neural control for state and output feedback: A novel formulation for efficient synthesis and verification, 2024.

\bibitem[Zhang et~al.(2023)Zhang, Xiu, Qu, and Fan]{pmlr-v211-zhang23a}
Zhang, S., Xiu, Y., Qu, G., and Fan, C.
\newblock Compositional neural certificates for networked dynamical systems.
\newblock In Matni, N., Morari, M., and Pappas, G.~J. (eds.), \emph{Proceedings of The 5th Annual Learning for Dynamics and Control Conference}, volume 211 of \emph{Proceedings of Machine Learning Research}, pp.\  272--285. PMLR, 15--16 Jun 2023.

\bibitem[Zhou et~al.(2022)Zhou, Quartz, Sterck, and Liu]{zhou2022neural}
Zhou, R., Quartz, T., Sterck, H.~D., and Liu, J.
\newblock Neural lyapunov control of unknown nonlinear systems with stability guarantees.
\newblock In \emph{Advances in Neural Information Processing Systems}, 2022.

\end{thebibliography}
\bibliographystyle{icml2024}

\newpage
\appendix
\onecolumn
\section{Code Availability}
The code for our proposed algorithm CoNSAL is available at \url{https://github.com/HaohanZou/CoNSAL}.

\section{Dynamics of the Tested Systems}\label{apdx:dynamics}
\subsection{\textbf{Inverted Pendulum}}
The inverted pendulum is a well-known classical nonlinear system that contains two state variables. The dynamics are formulated as follows,
\begin{align*}
    \Dot{x}_1 &= x_2, \\
    \Dot{x}_2 &= - \frac{g}{l} sin(x_1) - \frac{b}{m}x_2,
\end{align*}
where $x_1$ is the angular position from the inverted position, $x_2$ is the angular velocity, and parameters $g, m, l, b$ are acceleration of gravity, the mass of the inverted object, the length of string, and the coefficient of friction respectively. In experiment, setting $g = 9.81$, $m = \SI{2}{\kilo\gram}$, $l = \SI{5}{\metre}$, and $b = 0.1$, our proposed method finds the valid Lyapunov function $V = 1 - cos(x_1) + 0.254x_{2}^2$ over the state space: $\mathcal{D} = \{ (x_1, x_2) \in \mathbb{R}^2 \bm{\mid} |x_1| \leq \pi \text{ and } |x_2| \leq 6 \}$. This found Lyapunov function has the same analytical structure as the energy function of the inverted pendulum. 

\subsection{\textbf{Van Der Pol Oscillator}}\label{sub:vanderpol}
Van Der Pol Oscillator is a nonconservative, oscillating system with nonlinear damping \cite{zhou2022neural}. The dynamics of the Van Der Pol Oscillator have two state variables and are formulated as follows,
\begin{align*}
    \Dot{x}_1 &= x_2, \\
    \Dot{x}_2 &= -x_1 - \mu(1 - x_{1}^2) \cdot x_2,
\end{align*}
where $x_1$ and $x_2$ represent the object's position in the Cartesian coordinate, parameter $\mu \in \mathbb{R}^{+}$ indicates the strength of the damping. Under the state space $\mathcal{D} = \{ (x_1, x_2) \in \mathbb{R}^2 \bm{\mid} |x_i| \leq 1\}$ and setting $\mu = 1$, our proposed method found valid local Lyapunov function $V(x_1, x_2) = x_{1}^2 + x_{2}^2$. 
Other forms of Lyapunov functions for Van Der Pol Oscillator, for example, $V(x_1, x_2) =(x_1 \cdot (x_1+x_2) + x_{2}^2)/(2.8 - x_1 - x_2)$, are also recovered during the experiments.

\subsection{3-D Trig Dynamics}
3-D trig dynamics comes from exercise problems from textbook \cite{khalil2002nonlinear}, the  dynamics are written as follows,
\begin{align*}
    \Dot{x}_1 &= x_2, \\
    \Dot{x}_2 &= -2h(x_1) - x_2 - 2h(x_3), \\
    \Dot{x_3} &= x_2 - x_3,
\end{align*}
where $h(x) = sin(x)cos(x)$. When the state space is $\mathcal{D} = \{ (x_1, x_2, x_3) \in \mathbb{R}^3 \bm{\mid} |x_i| \leq 1.5, \forall i \in\{1,2,3\} \}$, the valid Lyapunov function found by our proposed method is $V(x_1, x_2, x_3) = 2 sin(x_1) + x_{2}^2 + 2 sin(x_3)^2$, which is consistent to the textbook solution of Lyapunov function for this particular dynamics.

\subsection{\textbf{Rotating Wheel Pendulum}} \label{sub:Rotating Wheel Pendulum}
The rotating wheel pendulum is structured as a pendulum with an additional rotating wheel at the end string, which freely spins along an axis parallel to the axis of the pendulum \cite{wheel-pendulum}. The dynamics of the system have four state variables and are formulated as follows,

\begin{align*}
    \Dot{x}_1 &= x_2, \\
    \Dot{x}_2 &= \frac{d_{22}}{det(D)} \bar{m}g \cdot x_1 + \frac{-d_{12}}{det(D)} \cdot \tau, \\
    \Dot{x}_3 &= x_4, \\
    \Dot{x}_4 &= \frac{d_{21}}{det(D)} \bar{m}g \cdot x_1 + \frac{d_{11}}{det(D)} \cdot \tau ,\\
    \tau &= \frac{-x_4 - x_3 + k_1 sin(x_1)}{E + k_2},
\end{align*}
where $x_1$ is the angular position from the inverted position of the pendulum, $x_2$ is the angular velocity of the pendulum, $x_3$ is the angular position of the wheel, $x_4$ is the angular velocity of the wheel, $\tau$ is the torque applied to disk from the motor, $m_1$ is the mass of the pendulum, $m_2$ is the mass of the wheel, $l_1$ is the length of the pendulum, $l_2$ is the distance to the center of mass of the pendulum, $I_1$ is the moment of inertia of the pendulum, $I_2$ is the moment of inertia of the wheel, $\bar{m} = m_1l_2 + m_2l_1$, $k_1 = \frac{d_{21} \bar{m}g}{det(D)}$, and $k_2 = \frac{d_{11}}{det(D)}$. $E = \frac{1}{2} \begin{bmatrix}
    x_2&x_4 \\
\end{bmatrix} D \begin{bmatrix}
    x_2\\x_4
\end{bmatrix} + \bar{m}g(cos(x_1) - 1)$ is the energy function of the dynamics, where $D$ is the inertia matrix formulated as 
$$D = 
\begin{bmatrix}
 m_1 l_{2}^{2} + m_2 l_{1}^{2} + I_1 + I_2 & I_2\\
 I_2 & I_2
\end{bmatrix}.$$
We consider the state space as $\mathcal{D} = \{ (x_1, x_2, x_3, x_4) \in \mathbb{R}^4 \bm{\mid} |x_1| \leq \frac{\pi}{2},|x_3| \leq \frac{\pi}{2}, |x_2| \leq 2, \text{ and } |x_4| \leq 2 \}$.

With control law $\tau = \frac{-x_4 - x_3 + k_1 sin(x_1)}{E + k_2}$, the system is stable. For this system, directly learning a Lyapunov function is difficult. Thus, we investigate the benefit of using prior knowledge of the system for constructing the Lyapunov function. 
Specifically, we use the energy function $E$ as the additional information on the dynamics and use it as an additional variable to the neural Lyapunov function during training.

We set $m_1 = (\frac{0.1}{9.81}) \si{\kilo\gram}$, $m_2 = (\frac{0.4}{9.81})\si{\kilo\gram}$, $l_1 = l_2 =$ \SI{1}{\metre}, $I_1 = (1 - \frac{0.5}{9.81}) \si{\kilo\gram\cdot\metre\squared}$, and $I_2 = 1\si{\kilo\gram\cdot\metre\squared}$, the proposed method found a valid Lyapunov function $V(x_1, x_2, x_3, x_4, E) = E^2 + x_{3}^2 + x_{4}^{2}$.

\subsection{6-D Nonlinear Dynamics}

This high-dimensional dynamics consists of three two-dimensional asymptotically stable linear subsystems that are coupled by three nonlinearities with small gains adopted from \cite{6ddynamics}. The dynamics are written as 
\begin{align*}
    \Dot{x}_1 &= - x_1 + 0.5 x_2 - 0.1 x_{5}^{2}, \\
    \Dot{x}_2 &= -0.5 x_1 - x_2, \\
    \Dot{x}_3 &= - x_3 + 0.5 x_4 - 0.1 x_{1}^{2}, \\
    \Dot{x}_4 &= -0.5 x_3 - x_4, \\
    \Dot{x}_5 &= - x_5 + 0.5 x_6, \\
    \Dot{x}_6 &= -0.5 x_5 - x_6 + 0.1 x_{2}^{2}.
\end{align*}
    
Our proposed method is able to find a valid Lyapunov function $V(x) = 4x_{1}^{2} - 1.24x_{2}^{4} + (x_1 + x_2 + x_3 + x_4 + x_5)^2 + 2.996 \cdot (0.578x_{2}^{2} + 0.578x_{3}^{2} + 0.578x_{5}^{2} + 0.578x_{6}^{2} + 1)^2$ over the region $\mathcal{D} = \{ (x_1, x_2, x_3, x_4, x_5, x_6) \in \mathbb{R}^6 \bm{\mid} |x_i| \leq 1, \forall i \in \{1,2,..,6\}\}$, which is a valid Lyapunov function for the given dynamics.

\subsection{\textbf{Power System}}
We test the compositional neural Lyapunov function design with a 3-bus power system \cite{cuipower} to examine its ability on high-dimensional dynamics. Consider $\theta_i$, $\omega_i$ as the phase angle and the frequency of bus $i$, respectively, the dynamics for each bus are formulated as follows,
\begin{align*}
   \Dot{\theta_{i}} &= \omega_{i}, \\
    m_{i} \Dot{\omega} &= p_{i} - d_i \omega_i - u_i(\omega_i) - \sum_{j = 1}^{3} B_{ij} sin(\theta_i - \theta_j),
\end{align*}
where $m_i$ is the generator inertia constant, $d_i$ is the combined frequency response coefficient from synchronous
generators and frequency sensitive load, and $p_i$ is the net power injection, for each bus $i = 1, 2, 3$. $B \in \mathbb{R}^{3 \times 3}$ is the susceptance matrix with $B_{ij} = 0 $ for every pair $\{i, j\}$ such that bus $i$ and bus $j$ are not connected, and $u_i(\omega_i)$ is the controller at bus $i$ that adjusts the power injection to stabilize the frequency.

Since the frequency dynamics of the system depends
only on the phase angle differences, so we change the coordinates:
$$\delta_i = \theta_i - \frac{1}{3}\sum_{i = 1}^{3} \theta_i$$
where $\delta_i$ can be understood as the center-of-inertia coordinates of each bus. In our experiment, for simplicity, we set $p_i = 0$, $m_i = 1$, $d_i = 1$, $u_i(\omega_i) = \omega_i$, and $B_{ij} = 1 \: \forall \: i \neq j, B_{ii} = 0$. In this case, the equilibrium point for our system is at the origin, i.e., $\delta_i^{*} =\omega_i^*= 0, \: i = 1, 2, 3$. The state space for our experiment is defined as: $\mathcal{D} = \{ (\delta_1,\delta_2,\delta_3, \omega_1,\omega_2,\omega_3) \in \mathbb{R}^6 \bm{\mid} |\delta_i| \leq \frac{\pi}{4} \text{ and } |\omega_i| \leq 2 \text{ for } i = 1, 2, 3\}$.

For the training procedures, we composed and jointly trained two neural networks for these dynamics. First, for each distinct bus i, $(\delta_i, \omega_i) \in \mathbb{R}^2$ are fed into an Input Convex Neural Network (ICNN). Then, for each pair of connected buses $\{i,j\}$, we input their frequency differences $\delta_i - \delta_j$ into another feed-forward Neural Network. In the end, symbolic regression is applied to each distinct neural network, and found Lyapunov function is the composition of regression formulas from two neural networks.

Through our method,
we retrieved a valid Lyapunov function $V(\delta_1, \delta_2, \delta_3, \omega_1, \omega_2, \omega_3) = 0.516  \left(\sum\limits_{i=1}^{3} \omega_{i}^2 \right) -0.5 \sum\limits_{i=1}^{3} \sum\limits_{\substack{j=1 \\ i\neq j}}^{3}\left(cos(\delta_i - \delta_j) - 1 \right)$ from the joint training of two neural networks.

\section{Neural Network Structure and Symbolic Regression Setting}\label{apdx:neural}
We now detail the neural network structures and the base operators used for symbolic regression in the following table. For simplicity, we set $F(\cdot)$ defined in \eqref{eq:NN-Ly} as $F(x)=x$.
\begin{table}[h]\label{table: Neural Network Structure}
\centering
\begin{tabular}{ |p{5.4cm}||p{4.5cm}|p{4.5cm}| }
    \hline
    \textbf{Dynamics} & \textbf{Neural Lyapounov Function} & \textbf{Symbolic Regression Unary Operator}  \\
    \hline
    \hline
    2-D Path Following (Linear) & (2, 128, 128, 1) & [$sin$, $cos$, $1-cos$, $x^2$] \\
    \hline
    2-D Inverted Pendulum  & (2, 128, 128, 1) & [$sin$, $cos$, $1-cos$, $x^2$]  \\
    \hline
    2-D Van Der Pol Oscillator&   (2, 128, 128, 1)  & [$sin$, $cos$, $1-cos$, $x^2$] \\
    \hline
    3-D Trig Dynamics &   (3, 128, 128, 1)  & [$sin$, $cos$, $1-cos$, $x^2$]  \\
    \hline
    4-D Rotating Wheel Pendulum & (4, 128, 128, 1) & [$sin$, $cos$, $1-cos$, $x^2$] \\
    \hline
    6-D Nonlinear Dynamics & (6, 128, 128, 1) & [$x^2$] \\
    \hline
    6-D 3-bus Power System & (2, 128, 128, 1) \& (1, 32, 32, 1) & [$x^2$] \& [$x^2$, $cos$, $1-cos$] \\
    \hline
\end{tabular}
\caption{Network size and regression model set up for each task. The tuple denotes the number of neurons in each layer of Neural Network. The set denotes the pre-defined operators for Symbolic Regression model. For all experiments, we set binary operators as $[+, -, \times, \div]$.}
\end{table}




\end{document}